\documentclass[aps,twocolumn,showpacs,preprintnumbers,amsmath,amssymb,pre]{revtex4-1}
\usepackage{bm}
\usepackage{graphicx}
\newcommand{\D}{{\rm d}}

\begin{document}

\title{Unconventional entropy production in the presence of momentum-dependent forces}

\author{Chulan~Kwon$^{1}$}
\author{Joonhyun Yeo$^{2}$}
\author{Hyun Keun Lee$^{3}$}
\author{Hyunggyu Park$^4$}
\address{$^1$ Department of Physics, Myongji University, Yongin, Gyeonggi-Do 449-728, Korea\\
$^{2}$ Department of Physics, Konkuk University, Seoul 143-701, Korea\\
$^{3}$ Department of Physics and Astronomy, Seoul National University, Seoul 151-742, Korea\\
$^4$ School of Physics and QUC, Korea Institute for Advanced Study, Seoul 130-722, Korea}

\date{\today}

%\author{Chulan~Kwon$^{1}$, Joonhyun Yeo$^{2}$, Hyun Keun Lee$^{3}$ and Hyunggyu Park$^{4}$}
%\address{$^{1}$ Department of Physics, Myongji University, Yongin, Gyeonggi-Do 449-728, Korea}
%\address{$^{2}$ Department of Physics, Konkuk University, Seoul 143-701, Korea}
%\address{$^{3}$ Department of Physics and Astronomy, Seoul National University, Seoul 151-742, Korea}
%\address{$^{4}$ School of Physics and QUC, Korea Institute for Advanced Study, Seoul 130-722, Korea}
%\date{\today}
%\pacs{05.70.Ln}{Nonequilibrium and irreversible thermodynamics}
%\pacs{02.50.-r}{Probability theory, stochastic processes, and statistics}
%\pacs{05.40.-a}{Fluctuation phenomena, random processese, noise, and Brownian motion}

\begin{abstract}
We investigate an unconventional nature of the entropy production (EP) in nonequilibrium systems with 
odd-parity variables that change signs under time reversal. 
We consider the Brownian motion of a particle in contact with a heat reservoir, where particle
momentum is an odd-parity variable. In the presence of an {\it external} momentum-dependent force,
the EP transferred to environment is found {\em not} equivalent to
usual reservoir entropy change due to heat transfer.  There appears an additional unconventional contribution to the EP,
which is crucial for maintaining the non-negativity of the (average) total EP enforced by the thermodynamic second law.
A few examples are considered to elucidate the novel nature of the EP. We also discuss detailed balance conditions 
with a momentum-dependent force.
\end{abstract}

%\submitto{\JSTAT}

%\keywords{Keywords: Stochastic particle dynamics (Theory), Fluctuations (Theory), Stochastic processes (Theory)}
% \noindent{\it Keywords\/}: Stochastic particle dynamics (Theory), Fluctuations (Theory), Stochastic processes (Theory)

\maketitle

\section{Introduction}
Recent studies on nonequilibrium (NEQ) fluctuation were motivated by the discovery of the fluctuation theorems (FT's)~\cite{evans,jarzynski,crooks,kurchan,lebowitz}.
The FT's were at first regarded as new identities or relations governing thermally fluctuating quantities in various NEQ dynamics, deterministic or
stochastic. More recent studies, however, have revealed many interesting properties beyond simple relations, which include excess and housekeeping
(or nonadiabatic and adiabatic) contributions to the entropy production (EP)~\cite{sasa,ge-hong,seifert,esposito,spinney,hklee},
modifications of FT~\cite{zon,puglish,jd,kmkim}, information entropy~\cite{sagawa,ito},
multiple dynamical transitions~\cite{jd-kwon-park}, and persistent initial-memory effect~\cite{farago,jslee}.

Most of the studies so far have dealt with stochastic systems where state variables have even parities (do not change signs) under time reversal.
This is mainly due to technical simplicity and might be also to a naive presumption that the generalization to the case including odd-parity variables might be straightforward.
However, it has been recently recognized that systems with odd-parity variables exhibit a fundamental difference from those with  even-parity variables only. For example, the housekeeping EP does not satisfy the FT any longer and can be divided into two parts characterized by the breakage of detailed balance (DB) and the parity asymmetry of the steady-state
probability distribution function~\cite{spinney,hklee}.

In this study, we report another fundamental difference associated with the nature of the EP in systems with both
even- and odd-parity variables from those without odd-parity variables.
To be specific, we consider an external momentum-dependent driving force in the underdamped Brownian motion.
Typical examples are velocity-dependent forces in active matter systems~\cite{active} such as
granular particles under vibration~\cite{granular}, interacting molecular motors~\cite{mm}, and active Brownian
particles~\cite{GC,active_Brown}. Simpler examples are the Lorentz force on a charged particle in a uniform magnetic field and
an additional dissipative force through a feedback process in a molecular refrigerator (cold damping)~\cite{mm,ito,cold_damping}.

For a system in contact with a single thermal reservoir at temperature $T$, the conventional
EP in the NEQ steady state can be characterized by an incessant environmental EP mediated solely by heat dissipation $Q$ into the
reservoir such that $\Delta S_{env}=\Delta S_{res}=Q/T$. However, we find that this relation should be modified in the presence of external momentum-dependent forces and there appears an extra environmental EP, $\Delta S_{uc}$ (unconventional EP), yielding $\Delta S_{env}=\Delta S_{res} + \Delta S_{uc}$.
In some cases, we can explicitly show that
$\Delta S_{res}$ is negative by itself, and the positivity of  $\Delta S_{env}$ is restored
by adding $\Delta S_{uc}$.
Therefore, the existence of $\Delta S_{uc}$ is crucial for the validity of the thermodynamic second law  ($\Delta S_{env} \ge 0$ in the steady state)~\cite{park_talk}.
A similar extra contribution to the environmental EP has been considered in Ref.~\cite{GC}
for a one dimensional system with a simple form of the momentum-dependent force.

%\vspace{3mm}
\section{Model}
Consider the driven Brownian motion of a particle of mass $m$
in state space $\mathbf{q}=(\mathbf{x},\mathbf{p})$, where $\mathbf{x}$ and $\mathbf{p}$ are the position and momentum vectors
in $d$ space dimensions, respectively.
The stochastic differential equation reads
\begin{equation}
\label{SDE}
\frac{\D\mathbf{x}}{\D t} = \frac{\mathbf{p}}{m}~, \quad
\frac{\D\mathbf{p}}{\D t} =\mathbf{f}(\mathbf{q};\lambda(t))
-\mathsf{G}\cdot\frac{\mathbf{p}}{m}+\boldsymbol{\xi}(t)~,
\end{equation}
where $\mathsf{G}$ is a dissipation matrix and a dot ($\cdot$) between matrices
(or vectors) denotes a contraction (inner product).
The Gaussian noise vector $\boldsymbol{\xi}(t)$ satisfies
$\langle \boldsymbol{\xi}(t) \boldsymbol{\xi}^\text{t} (t')\rangle=2
\mathsf{D}\delta(t-t')$ with superscript `$\text{t}$' denoting `transpose'. The diffusion matrix $\mathsf{D}$ is positive definite and symmetric.
The last two terms of the above equation describe {\em thermal forces} exerted by heat reservoirs.
With a single heat reservoir at temperature $T$, the Einstein relation holds as $\mathsf{D}= T\mathsf{G}$.
The most general form of the force $\mathbf{f}$ is considered,
which may depend on both position and momentum as well as on a time-dependent protocol $\lambda(t)$.
We note that previous studies \cite{GC} have only focused on the case where $\mathbf{f}$
is given by the sum of two terms, one of which depends only on position and the other on momentum.

The corresponding Kramers equation for the probability distribution function (PDF), $\rho(\mathbf{q},t)$, reads
\begin{equation}
\partial_t \rho=-\mathsf{L}\rho~,
\label{kramers}
\end{equation}
with the evolution operator $\mathsf{L}$ as
\begin{equation}
\mathsf{L}(\mathbf{q})= \partial_{\mathbf{x}} \cdot\frac{\mathbf{p}}{m}
+\partial_{\mathbf{p}}\cdot\left(\mathbf{f}-\mathsf{G}\cdot\frac{\mathbf{p}}{m}-\mathsf{D}\cdot\partial_{\mathbf{p}}\right)~.
\label{Hamiltonian}
\end{equation}
The formal solution can be written in terms of a path integral for $0\le t\le\tau$ as
\begin{equation}
\rho(\mathbf{q}, \tau)=\int \D\mathbf{q}_0 \int D [\mathbf{q}(t)] ~\Pi[\mathbf{q}(t);\lambda(t)]~ \rho(\mathbf{q}_0)~,
\end{equation}
with the initial PDF $\rho(\mathbf{q}_0)$. The kernel
$\Pi[\mathbf{q}(t);\lambda(t)]$ is the conditional path probability
density of the system evolving along a given path $\{\mathbf{q}(t)\}$ for $0\le t\le\tau$, starting at
$\mathbf{q}(0)=\mathbf{q}_0$.
The integration $\int D [\mathbf{q}(t)]$ is over all paths for given $\mathbf{q}_0$ and $\mathbf{q(\tau)}$.
The probability density to find a path $\{\mathbf{q}(t)\}$  is given
by $P[\mathbf{q}(t);\lambda(t)]=\Pi[\mathbf{q}(t);\lambda(t)]~\rho(\mathbf{q}_0)$.

%\vspace{3mm}
\section{Irreversibility}
Entropy production measures irreversibility of the process. To define irreversibility, we first consider
the time-reverse process which should be governed by the equivalent equation of motion to
that of the original (time-forward) process in Eq.~(\ref{SDE}):
\begin{equation}
\label{SDEr}
\frac{\D\mathbf{\bar{x}}}{\D \bar{t}} = \frac{\mathbf{\bar{p}}}{m}~, \quad
\frac{\D\mathbf{\bar{p}}}{\D \bar{t}} =\mathbf{f}(\mathbf{\bar{q}};\bar{\lambda}(\bar{t}))
-\mathsf{G}\cdot\frac{\mathbf{\bar{p}}}{m}+\boldsymbol{\xi}(\bar{t})~,
\end{equation}
where all variables in the time-reverse process are denoted by overbars. The time-dependent protocol function
$\bar{\lambda}$ is set by requiring $\bar{\lambda}(\bar{t})=\lambda(t)$.

For each (time-forward) path $\{\mathbf{q}(t)\}$ ($0\le t\le\tau$), one can consider the
corresponding time-reverse path $\{\bar{\mathbf{q}}(\bar{t})\}$ for $0\le \bar{t}\le\tau$ with $\bar{t}=\tau-t$
and $\bar{\mathbf{q}}(\bar{t}) =\varepsilon\mathbf{q}(t)$,
where parity $\varepsilon=1$ for position $\mathbf{x}$ (even parity)
and $\varepsilon=-1$ for momentum $\mathbf{p}$ (odd parity).
Note that the odd parity for momentum is automatically enforced by identifying $\bar{\mathbf{x}}(\bar{t})=\mathbf{x}(t)$
with $\bar{t}=\tau-t$ in Eq.~(\ref{SDEr}).
Irreversibility (or the total EP, $\Delta S_{tot}$) for a given path $\{\mathbf{q}(t)\}$ is defined by
\begin{equation}
\Delta S_{tot}[\mathbf{q}(t)]=
\ln \frac{\Pi[\mathbf{q}(t);\lambda(t)]\rho(\mathbf{q}_0)}
{\bar{\Pi}[\bar{\mathbf{q}}(\bar{t});\bar{\lambda}(\bar{t})]\bar\rho(\mathbf{\bar {q}}_0)}
=\ln \frac{P[\mathbf{q}(t);\lambda(t)]}{\bar{P}[\bar{\mathbf{q}}(\bar{t});\bar{\lambda}(\bar{t})]}~,
\label{tot_entropy_ratio}
\end{equation}
where $\bar{\Pi} [\bar{\mathbf{q}}(\bar{t});\bar{\lambda}(\bar{t})]$ is the conditional path probability density for the time-reverse path $\{\bar{\mathbf{q}}(\bar{t})\}$ and its initial PDF is given as $\bar\rho(\mathbf{\bar {q}}_0)=\rho(\mathbf{{q}}(\tau))$.

For a {\em reversible} path, path probabilities $P$ and ${\bar P}$
are the same by definition and we expect no EP. Deviation from this accounts for irreversibility of the path.
The usual (average) environmental EP, $\langle \Delta S_{env}\rangle$, is obtained by averaging
over all possible paths and the initial PDF.
The usual (average) total EP, $\langle \Delta S_{tot}\rangle$, is obtained by averaging
over all possible paths.
A simple probability normalization condition results in the integral FT,
expressed as $\langle e^{-\Delta S_{tot}}\rangle=1$~\cite{seifert,esposito}.
With the Jensen's inequality, it leads to $\langle \Delta S_{tot}\rangle\ge 0$, which proves the second law of thermodynamics.

The total EP can be divided into the system entropy change and the environmental EP as $\Delta S_{tot}=\Delta S_{sys}+\Delta S_{env}$,
where $\Delta S_{sys}=-\ln \rho(\mathbf{q}(\tau))+\ln\rho(\mathbf{q}_0)$
is given by Shannon entropy difference. Thus, from Eq.~(\ref{tot_entropy_ratio}), the environmental EP $\Delta S_{env}[\mathbf{q}(t)]$
is simply the logarithm of two conditional path probability density ratio.

It is useful to discretize time such that $t_i=i\Delta t$ for $i=0,\ldots, N$
with interval $\Delta t=\tau/N$ for large  $N$. A path is now represented by a discrete sequence $\{\mathbf{q}(t)\}=\{\mathbf{q}_0,\ldots,\mathbf{q}_i,\ldots,\mathbf{q}_N\}$
where $\mathbf{q}_i=\mathbf{q}(t_i)$. The Markovian property of dynamics allows us to write
\begin{equation}
\Pi[\mathbf{q}(t);\lambda(t)]=\prod_{i=1}^{N}\Gamma(\mathbf{q}_i|\mathbf{q}_{i-1}; \lambda_{i-1})
\end{equation}
where $\Gamma(\mathbf{q}_i|\mathbf{q}_{i-1}; \lambda_{i-1})
=\langle\mathbf{q}_i|e^{-\Delta t \mathsf{L}}|\mathbf{q}_{i-1}\rangle$ is the conditional probability
for the system at $(\mathbf{q}_i, t_i)$, starting from $(\mathbf{q}_{i-1}, t_{i-1})$
with $\lambda_{i-1}=\lambda(t_{i-1})$.
Using a quantum mechanical formalism for the non-Hermitian evolution operator $\mathsf{L}$, we find
\begin{eqnarray}
\lefteqn{\Gamma(\mathbf{q}_i|\mathbf{q}_{i-1}; \lambda_{i-1})= \frac{\delta(\mathbf{x}_i-\mathbf{x}_{i-1}-\Delta t\;\mathbf{p}_i^{(\alpha)}/m)}{|\det(4\pi\Delta t \mathsf{D})|^{d/2}} }\label{trans_prob}\\
&& \times \exp\left\{-\frac{\Delta t}{4} \mathbf{h}_i^{(\alpha)}\cdot \mathsf{D}^{-1}\cdot \mathbf{h}_i^{(\alpha)}+\alpha\Delta t\left(\frac{\textrm{Tr}\mathsf{G}}{m}-\partial_{\mathbf{p}^{(\alpha)}}\cdot\mathbf{f}_i^{(\alpha)}\right)\right\}\nonumber
\end{eqnarray}
where
\begin{equation}
\mathbf{h}_i^{(\alpha)}=\frac{\mathbf{p}_i-\mathbf{p}_{i-1}}{\Delta t}+\mathsf{G}\cdot\frac{\mathbf{p}_i^{(\alpha)}}{m}-\mathbf{f}_i^{(\alpha)}~,
\end{equation}
and $\mathbf{p}_i^{(\alpha)}=(1-\alpha)\mathbf{p}_{i-1}+\alpha\mathbf{p}_i$ for $0\le \alpha \le 1$ represents an intermediate value
of $\mathbf{p}$ during time interval $[t_{i-1}, t_{i}]$. In principle, $\alpha$ can be arbitrary, but should
not affect on physical observables like EPs for the large $N$ (small $\Delta t$) limit. Force
$\mathbf{f}_i^{(\alpha)}$ is also defined in the same manner.

It is more convenient to consider the EP during an infinitesimal time interval $[t, t+\D t]$ as
\begin{equation}
\D S_{env}=\ln \frac{\Gamma(\mathbf{q}^\prime, t+\D t|\mathbf{q}, t;\lambda(t))}
{\Gamma(\varepsilon\mathbf{q}, t+\D t|\varepsilon\mathbf{q}^\prime, t;\lambda(t))}~,
\label{env_entropy_inf}
\end{equation}
where the conditional probability for the time-reverse path ($\mathbf{\bar{q}^\prime}\rightarrow\mathbf{\bar{q}}$) is
expressed in terms of that for the time-forward path ($\varepsilon\mathbf{{q}}^\prime\rightarrow\varepsilon\mathbf{{q}}$)
in the denominator.
Then, a straightforward algebra yields
\begin{equation}
\frac{\D S_{env}}{\D t}=
-\left(\frac{\D\mathbf{p}}{\D t}-{\mathbf{f}^{\rm rev}}\right)\cdot \mathsf{D}^{-1}\cdot \left(
\mathsf{G}\cdot\frac{\mathbf{p}}{m} -\mathbf{f}^{\rm ir}\right) - \partial_{\mathbf{p}}\cdot\mathbf{f}^{\rm rev}~,
\label{env_entropy}
\end{equation}
where the force is divided into the reversible and irreversible parts,
$\mathbf{f}(\mathbf{q})=\mathbf{f}^{\rm rev} (\mathbf{q})+\mathbf{f}^{\rm ir}(\mathbf{q})$, such that
$\mathbf{f}^{\rm rev} (\mathbf{q})=\mathbf{f}^{\rm rev}(\varepsilon\mathbf{q})$
and $\mathbf{f}^{\rm ir}(\mathbf{q})=-\mathbf{f}^{\rm ir}(\varepsilon\mathbf{q})$~\cite{risken}.
All quantities in the above equation should be evaluated at the midpoint ($\alpha=1/2$; Stratonovich convention).
We note that the resulting expression is unique with midpoint values, independent
of choices of discretization schemes (i.e.~any $\alpha$ for the time-forward path and any $\bar\alpha$ for the time-reverse
path)~\cite{private_kwon}.

%\vspace{3mm}
\section{Currents}
It is illuminating to consider the average EP in terms of currents $\mathbf{j}_{\mathbf{q}}$ defined by the
continuity equation as
\begin{equation}
\partial_t \rho = -\partial_{\mathbf{q}}\cdot \mathbf{j}_{\mathbf{q}}=-\mathsf{L}\rho~,
\label{currents}
\end{equation}
where $\mathbf{j}_{\mathbf{q}}$  can be obtained from
the evolution operator $\mathsf{L}$ in Eq.~(\ref{Hamiltonian}).
We also divide the currents into the reversible and irreversible parts~\cite{risken}  as
\begin{eqnarray}
&&\mathbf{j}^{\rm rev}_\mathbf{p} =\mathbf{f}^{\rm rev}(\mathbf{q})\rho(\mathbf{q}), \label{j_rev}\\
&&\mathbf{j}^{\rm ir}_\mathbf{p} = \left( \mathbf{f}^{\rm ir}(\mathbf{q})
-\mathsf{G}\cdot\frac{\mathbf{p}}{m}
-\mathsf{D}\cdot\partial_\mathbf{p} \right)\rho(\mathbf{q})~, \label{j_ir}
\end{eqnarray}
with
\begin{equation}
\mathbf{j}^{\rm rev}_\mathbf{x} = \frac{\mathbf{p}}{m} \rho(\mathbf{q}) \qquad \text{and}\qquad
\mathbf{j}^{\rm ir}_\mathbf{x} = \mathbf{0}~.
\label{current_position}
\end{equation}
It is trivial to show that
\begin{equation}
\left\langle \frac{\D S_{sys}}{\D t}\right\rangle=
-\frac{\D}{\D t}\int \D \mathbf{q}~ \rho \ln \rho=
-\int \D \mathbf{q}~
\frac{\mathbf{j}_\mathbf{p}\cdot\partial_\mathbf{p}\rho }{\rho}~.
\label{sys_entropy}
\end{equation}
Averaging the environmental EP in Eq.~(\ref{env_entropy}), we need to evaluate two-time correlators at time $t$ and $t+\D t$, such as
$\langle A(\mathbf{q}^\prime)B(\mathbf{q}) \rangle =\int \D\mathbf{q}^\prime \D\mathbf{q}~
A(\mathbf{q}^\prime)\Gamma(\mathbf{q}^\prime,t+\D t|\mathbf{q},t) B(\mathbf{q}) \rho(\mathbf{q})$.
After some algebra, we find the average total EP rate as
\begin{equation}
\left\langle \frac{\D S_{tot}}{\D t}\right\rangle=
\int \D \mathbf{q}~
\frac{\mathbf{j}_\mathbf{p}^{\rm ir}\cdot \mathsf{D^{-1}} \cdot\mathbf{j}_\mathbf{p}^{\rm ir}}
{\rho}~\ge 0,
\label{tot_entropy}
\end{equation}
where the inequality (thermodynamic second law) comes from the positive-definiteness of
$\mathsf{D}$~\cite{exp1}. Only the irreversible part $\mathbf{j}_\mathbf{p}^{\rm ir}$ of the current is a single measure for the irreversibility.
In the overdamped limit, $\langle\mathbf{j}_\mathbf{x}^{\rm rev}\rangle_\mathbf{p}$ averaged over $\mathbf{p}$ plays an equivalent role, recognized as nonequilibrium
steady-state current in position space~\cite{seifert,kwon-ao-thouless}. Interestingly, $\langle\mathbf{j}_\mathbf{x}^{\rm rev}\rangle_\mathbf{p}\neq\mathbf{0}$
is not a necessary condition for irreversibility, which will be observed from the second example later.

%\vspace{3mm}
\section{Detailed balance}
Stochastic reversibility of a given process implies its time-reversal symmetry in the probabilistic sense,
which defines {\em equilibrium}. This is also known to be enforced by the detailed balance condition (DB)
between all pairs of microscopic states~\cite{risken,gardiner}.
However, when odd-parity variables are present, a special care is necessary to relate the DB condition to stochastic reversibility.
The DB condition in general
is given in terms of transition rates between $\mathbf{q}$ and $\mathbf{q}'$($\neq \mathbf{q}$) as
\begin{equation}
\langle \mathbf{q}'| \mathsf{L} | \mathbf{q}\rangle \rho(\mathbf{q})
=\langle \varepsilon\mathbf{q}| \mathsf{L} | \varepsilon\mathbf{q}'\rangle\rho(\varepsilon\mathbf{q}').
\label{DB}
\end{equation}
Conventionally,
the mirror symmetry of the steady-state PDF, $\rho_{ss}(\mathbf{q})=\rho_{ss}(\varepsilon\mathbf{q})$
is also required for the system to be in equilibrium.
Recently, it has been reported for {\em discrete} state jumping dynamics~\cite{hklee} that the DB condition does not necessarily imply
the mirror symmetry and that the two conditions are in fact independent.

For the Brownian motion considered here, we analyze the DB equation (\ref{DB}) without assuming the mirror symmetry beforehand.
In terms of the differential operator $\mathsf{L}(\mathbf{q})$ given in Eq.~(\ref{Hamiltonian}), the DB condition is
$
\mathsf{L}(\mathbf{q}')\delta(\mathbf{q}'-\mathbf{q})\rho(\mathbf{q})=
\mathsf{L}(\varepsilon\mathbf{q})\delta(\mathbf{q}-\mathbf{q}')\rho(\varepsilon\mathbf{q}').
$
Changing $\rho(\mathbf{q})$ to $\rho(\mathbf{q}')$ on the left-hand side and
moving $\rho(\varepsilon\mathbf{q}')$ on the right-hand side to the front, we can rewrite this as
\begin{equation}
\mathsf{L}(\mathbf{q}')\rho(\mathbf{q}')\delta(\mathbf{q}'-\mathbf{q})
=\rho(\varepsilon\mathbf{q}')\mathsf{L}^\dag(\varepsilon\mathbf{q}')\delta(\mathbf{q}'-\mathbf{q}),
\label{DB1}
\end{equation}
where the operator $\mathsf{L}^\dag$ is defined as
\begin{equation}
\mathsf{L}^\dag(\mathbf{q})\equiv - \frac{\mathbf{p}}{m}\cdot\partial_{\mathbf{x}}
-\left(\mathbf{f}-\frac{\mathbf{p}}{m}\cdot\mathsf{G}^{\text{t}}+\partial_{\mathbf{p}}\cdot\mathsf{D}\right)\cdot\partial_{\mathbf{p}}
\end{equation}
Equation (\ref{DB1}) can now be regarded as the equality between the differential operators acting on delta functions.
From the part that does not involve any derivatives on the delta function,
the steady state condition $\mathsf{L}(\mathbf{q}')\rho(\mathbf{q}')=0$ follows, so $\rho(\mathbf{q}')=\rho_{ss}(\mathbf{q}')$
and $\partial_{\mathbf{q}}\cdot \mathbf{j}_{ss,\mathbf{q}}=0$ .
Comparing the terms proportional to
$\partial_{\mathbf{x}'}\delta(\mathbf{q}'-\mathbf{q})$, we get the mirror symmetry
\begin{equation}
\rho_{ss}(\mathbf{q})=\rho_{ss}(\varepsilon\mathbf{q}).
\label{parity_symm}
\end{equation}
Thus, the mirror symmetry is a direct consequence of the DB condition for the Brownian dynamics,
in contrast to discrete state jumping dynamics. For a more general continuous stochastic dynamics involving multiplicative noises, however, the situation is not so simple.
In that case, the mirror symmetry follows
from the DB only after assuming a certain condition for the multiplicative noise strengths. In general the two conditions remain independent~\cite{yeo}.
The broken mirror symmetry manifests the existence of nonzero average current
$\langle\mathbf{j}_\mathbf{x}^{\rm rev}\rangle_\mathbf{p}$ in position space, as seen from Eq.~(\ref{current_position}).

The terms proportional to $\partial_{\mathbf{p}'}\delta(\mathbf{q}'-\mathbf{q})$
gives the vanishing irreversible steady state current
\begin{equation}
\mathbf{j}^{\rm ir}_{ss,\mathbf{p}}=\left(\mathbf{f}^{\rm ir}(\mathbf{q})-\mathsf{G}\cdot\frac{\mathbf{p}}{m}-\mathsf{D}
\cdot\partial_{\mathbf{p}}\right)\rho_{ss}(\mathbf{q})=\mathbf{0}.
\end{equation}
This is consistent with the previous result in Eq.~(\ref{tot_entropy}) in  that
the DB condition characterizing equilibrium guarantees stochastic reversibility
even in the presence of momentum-dependent forces.
With the broken DB, the total EP should increase incessantly in time in NEQ steady states.
Higher-order derivative terms do not provide any additional condition.

%\vspace{3mm}
\section{Unconventional EP}
The definition of the environmental EP, Eqs.~(\ref{env_entropy_inf}) and (\ref{env_entropy}), in terms of the irreversibility measure should be checked against its conventional definition in thermodynamics. Without any momentum-dependence force,
$\mathbf{f}(\mathbf{{q}})=\mathbf{f}(\mathbf{{x}})=\mathbf{f}^{\rm rev}$ and $\mathbf{f}^{\rm ir}=0$, it has been already shown in literatures~\cite{seifert,esposito} that this definition is consistent with the conventional one. For example, in the case of a single heat reservoir at temperature $T$ with $\mathsf{D}= T\mathsf{G}$, Eq.~(\ref{env_entropy}) is simplified as
\begin{equation}
\frac{\D S_{env}}{\D t}=
-\frac{1}{T} \left(\frac{\D\mathbf{p}}{\D t}-{\mathbf{f}}\right)\cdot \frac{\mathbf{p}}{m}
=\frac{1}{T}\frac{\D Q}{\D t}~,
\label{env_entropy_simple}
\end{equation}
where $\D Q$ is the heat flow into the reservoir (minus work done by the thermal forces)
along a given trajectory during an infinitesimal time interval $\D t$.
(Note that heat should be evaluated with the midpoint value of $\mathbf{p}/m$ in the above equation
to maintain the energy conservation law~\cite{Chulan}). Thus,
the environmental EP is given solely by the conventional Clausius EP as
$\D S_{env}= \D S_{res}= \D Q /T$.
With general (multiple) reservoirs, we find
\begin{equation}
\frac{\D S_{env}}{\D t}=\frac{\D S_{res}}{\D t}=
-\left(\frac{\D\mathbf{p}}{\D t}-{\mathbf{f}}\right)\cdot
\mathsf{D}^{-1}\cdot\mathsf{G}\cdot\frac{\mathbf{p}}{m}~.
\label{env_entropy_multiple}
\end{equation}
With the temperature matrix $\mathsf{T}=\mathsf{G}^{-1}\cdot\mathsf{D}$ and heat matrix differential
$\D \mathsf{Q}=-(\D \mathbf{p} - \mathbf{f} \D t) (\mathbf{p}/m)^\text{t}$, we get
$\D S_{env}= \textrm{Tr} ({\mathsf{T}^{{-1}}} \cdot\D\mathsf{Q}^{\text t})$, which can be interpreted
as the generalized Clausius EP into reservoirs.

With momentum-dependent forces, it is obvious that the (generalized) Clausius EP rate is not identical to
the environmental EP: Eq.~(\ref{env_entropy}) can be rewritten as $\D S_{env}=\D S_{res} +\D S_{uc}$,
where the  unconventional EP rate is simply given by the difference between Eqs.~(\ref{env_entropy}) and (\ref{env_entropy_multiple}).
The unconventional EP may be interpreted as an additional EP into the external agent that
exerts momentum-dependent forces $\mathbf{f}(\mathbf{{q}})=\mathbf{f}(\mathbf{{x}},\mathbf{{p}})$.
However, this external agent is not in the conventional form such as a thermal reservoir, so
it is difficult to understand this additional entropy in terms of the conventional thermodynamics. In fact, we do not know
whether the unconventional EP has any feature of the conventional entropic measure, though it is evident that
it contributes to the irreversibility in Eq.~(\ref{tot_entropy_ratio}).
Note that the unconventional EP is present even when $\mathbf{f}^{\rm ir}=0$, if $\mathbf{f}^{\rm rev}$
includes a momentum dependence.

The average reservoir (generalized Clausius) EP rate can be obtained via a similar algebra used in deriving Eq.~(\ref{tot_entropy}) as
\begin{equation}
\left\langle \frac{\D S_{res}}{\D t}\right\rangle=
\int \D \mathbf{q}~
\frac{\mathbf{p}}{m}\cdot {\mathsf{G}^\text{t}}\cdot
\left(\mathsf{D}^{-1}\cdot\mathsf{G}\cdot\frac{\mathbf{p}}{m}+\partial_{\mathbf{p}}\right)\rho(\mathbf{q})~,
\label{res_entropy}
\end{equation}
and similarly we find
\begin{eqnarray}
\left\langle\frac{\D S_{uc}}{\D t}\right\rangle =
\int \D \mathbf{q}~\rho(\mathbf{q})&&\left[\mathbf{f}^{\rm ir}\cdot\mathsf{D}^{-1}\cdot 
\left(\mathbf{f}^{\rm ir}-2 \mathsf{G}\cdot \frac{\mathbf{p}}{m}\right)\right.\qquad\nonumber\\
&&\qquad\left.-\partial_{\mathbf{p}}\cdot\left(\mathbf{f}^{\rm rev}-\mathbf{f}^{\rm ir}\right)\right]~.
\label{an_entropy_average}
\end{eqnarray}
These two EP rates sum up into the average environmental EP rate which is
consistent with the difference between Eqs.~(\ref{tot_entropy}) and (\ref{sys_entropy}).

%\vspace{3mm}
\section{Examples}
We now demonstrate the novelty and importance of the unconventional EP in a few examples that may be realized in experiments.
First, we investigate a cold-damping problem with a dissipative external force,
$\mathbf{f}^{\rm ir}=-\mathsf{G}^\prime\cdot\mathbf{p}/m$
with a positive-definite $\mathsf{G}^\prime$. For simplicity, we take
$\mathbf{f}^{\rm rev}=0$, $\mathsf{G}=\gamma\mathsf{I}$, and
$\mathsf{D}=D\mathsf{I}$ in two dimensions with the identity matrix $\mathsf{I}$ and
the reservoir temperature $T=D/\gamma$.
The steady-state PDF, $\rho_{ss}(\mathbf{q})$, can be easily obtained in any linear diffusion system, shown in
Refs.~\cite{jd-kwon-park} and \cite{kwon-ao-thouless}. For given
$\mathsf{G}'=\gamma\left(
\begin{array}{cc}
r & a\\
b & r
\end{array}\right)$  with $a,~b>0$ and $r^2>ab$, a straightforward algebra with Eqs.~(\ref{tot_entropy}) and (\ref{res_entropy})
yields
\begin{eqnarray}
\label{negative_heat}
\left\langle\frac{\D S_{tot}}{\D t}\right\rangle&=& \frac{\gamma}{m} \frac{(a-b)^2}{2(1+r)}\\
\left\langle\frac{\D S_{res}}{\D t}\right\rangle&=& \frac{\gamma}{m}
\left[\frac{4(1+r)^2+(a-b)^2}{2(1+r)\left\{(1+r)^2-ab\right\}}-2\right]~.\nonumber
\end{eqnarray}
As the average system EP vanishes in the steady state,
$\langle\D S_{uc}/\D t\rangle$ is simply the difference between the above two EP rates.
We can see explicitly that the reservoir EP rate (heat production rate) can be negative for large $r$, but
the total EP rate remains positive (thermodynamic second law) due to the unconventional EP contribution.

In a cold-damping problem, heat should flow into the system
at a lower temperature from the higher-temperature reservoir in the steady state, which
dissipates by the additional dissipative external force. Thus, the reservoir entropy decreases incessantly,
which is compensated and usually overridden by the unconventional EP to maintain the thermodynamic second law. In a practical
cold-damping problem, a feedback mechanism through measurement of a particle momentum operates continuously
as this additional dissipative force, thus a usual information entropy (mutual information)~\cite{sagawa}
enters into the total EP in addition, which will be discussed elsewhere~\cite{Chulan_information}.

As a second example, we consider the case with $\mathbf{f}^{\rm ir}=-\mathsf{A}\cdot \mathbf{p}/m$
with an antisymmetric $\mathsf{A}=-\mathsf{A}^{\rm t}$.
Taking $\mathbf{f}^{\rm rev}= -\partial_{\mathbf{x}} V$ and $\mathsf{D}=T\mathsf{G}$,
this can describe a charged particle motion under a confined potential $V(\mathbf{x})$ with a
magnetic field, in contact with a single temperature reservoir.
From  Eqs.~(\ref{kramers}) and (\ref{Hamiltonian}), it is easy to show that $\rho_{ss}(\mathbf{q})\propto e^{-\frac{1}{T}\left\{\mathbf{p}^2/2m+V(\mathbf{x})\right\}}$,
which is the same Boltzmann distribution as for the equilibrium case without $\mathbf{f}^{\rm ir}$.
We find $\langle\D S_{res}/\D t\rangle=0$ (no heat production because $\mathbf{f}^{\rm ir}$ does not
work) from Eq.~(\ref{res_entropy}), but the DB is broken due to a nonzero
irreversible current in the momentum space as in Eq.~(\ref{j_ir}), leading to a nonzero total EP in the steady state in
Eq.~(\ref{tot_entropy}), as
\begin{equation}
\mathbf{j}_{ss,\mathbf{p}}^{\rm ir}=-\mathsf{A}\cdot\frac{\mathbf{p}}{m}\rho_{ss}~,
\left\langle\frac{\D S_{tot}}{\D t}\right\rangle= \frac{1}{m} \textrm{Tr}
\left[\mathsf{G}^{-1}\mathsf{A}\mathsf{A}^{\rm t}\right]~.
\label{2nd_example}
\end{equation}
Note that the unconventional EP is the only source for the total EP in this case.

For a charged particle motion in two dimensions with an external uniform magnetic field $\mathbf{B}$
applied in the perpendicular direction, the antisymmetric matrix can be written as
$\mathsf{A}=B\left(\begin{array}{cc} 0&-1\\1&0\end{array} \right)$. Then,
we get $\langle\D S_{tot}/\D t\rangle = 2B^2/(\gamma m)$ with $\mathsf{G}=\gamma\mathsf{I}$.
Due to the mirror symmetry in the steady state, there is no net current in position space, $\langle\mathbf{j}_\mathbf{x}^{\rm rev}\rangle_\mathbf{p}=\mathbf{0}$. For the system in a confined geometry, one can expect a boundary current,
which represents irreversibility in position space, even though there is no net bulk current. Irreversibility in momentum space implies the existence of a residual force persistent in thermal fluctuation, which can be observed in this example from the helicity describing the tendency of circulation around the external magnetic field. The helicity can be measured by $\langle\mathbf{p}(t+\Delta t)\times\mathbf{p}(t)\rangle/(\Delta t)$ for $\Delta t\to 0$
with the cross product $\times$, which can be found as
$\int \text{d}\mathbf{p}~\mathbf{j}_{ss,\mathbf{p}}^{\rm ir}\times\mathbf{p}=2T\mathbf{B}$. It signifies a difference from the equilibrium motion in the absence of a magnetic field. We remark that, in the case of an internally current-induced magnetic field
which should change its sign under time reversal~\cite{sakurai}, the magnetic force becomes reversible, thus
$\mathbf{f}^{\rm ir}=0$. Then, the DB is restored and the total EP is zero in the steady state.

In summary, we have considered the EP in a system containing both even and odd variables under time reversal. We have obtained
explicit expressions for the EP's and their average values. In the presence of an external momentum-dependent force, the
environmental EP contains not only the usual reservoir EP due to heat transfer, but also the unconventional EP. This additional
EP is crucial for the validity of the thermodynamic second law.

\begin{acknowledgments}
This research was supported by the NRF Grant No.~2013R1A1A2011079 (C.K.), 2014R1A1A2053362 (J.Y.),
2014R1A3A2069005 (H.K.L.), and 2013R1A1A2A10009722 (H.P.).
\end{acknowledgments}

%\section*{References}

\vfil\eject

\begin{thebibliography}{99}
\bibitem{exp1} A similar expression was reported for simpler cases in~\cite{seifert,spinney,ge,Tome}.
\end{thebibliography}
\end{document}